\documentclass[journal]{IEEEtran}
\usepackage{graphicx,amssymb,amsmath}
\usepackage{multicol}
\usepackage{CJK}
\usepackage[noadjust]{cite}
\usepackage{setspace}
\usepackage{stfloats}
\usepackage{midfloat}
\usepackage{flushend,cuted}
\usepackage{bm}
\usepackage{multirow}
\usepackage{array}
\usepackage{stfloats}
\usepackage{cite}
\usepackage{algorithm}
\usepackage{algpseudocode}
\usepackage{color}

\IEEEoverridecommandlockouts

\ifCLASSINFOpdf
\else
\fi
\newcommand{\E}{\textmd{E}}
\newcommand{\Var}{\textmd{Var}}
\newcommand{\diag}{\textmd{diag}}
\newcommand{\vect}{\textmd{vec}}

\allowdisplaybreaks
\begin{document}
\title{Downlink Achievable Rate Analysis in Massive MIMO Systems with One-Bit DACs}

\author{Yongzhi Li,
        Cheng Tao,
        A. Lee Swindlehurst,~\IEEEmembership{Fellow,~IEEE,}
        Amine Mezghani
        and~Liu Liu
\thanks{Y. Li, C. Tao, and L. Liu are with the Institute of Broadband
Wireless Mobile Communications, Beijing Jiaotong University, Beijing, 100044, China (email:
liyongzhi@bjtu.edu.cn; chtao@bjtu.edu.cn; liuliu@bjtu.edu.cn).}
\thanks{A. Mezghani and A. L. Swindlehurst are with the Center for Pervasive Communications and
Computing, University of California, Irvine, CA 92697 USA (e-mail: amezghan@uci.edu; swindle@uci.edu). A. L. Swindlehurst is also a Hans Fischer Senior Fellow of the Institute for Advanced Study at the Technical University of Munich.}
\thanks{The research was supported in part by Beijing Nova Programme (Grant No. xx2016023), the NSFC project under grant No.61471027, the Research Fund of National Mobile Communications Research Laboratory, Southeast University (No.2014D05, No. 2017D01), and Beijing Natural Science Foundation project under grant No.4152043. A.~Swindlehurst was supported by the National Science Foundation under Grant ECCS-1547155, and by the Technische Universit\"at M\"unchen Institute for Advanced Study, funded by the German Excellence Initiative and the European Union Seventh Framework Programme under grant agreement No. 291763, and by the European Union under the Marie Curie COFUND Program.}
}
\maketitle

\begin{abstract}
In this letter, we investigate the downlink performance of massive multiple-input multiple-output (MIMO) {\color{black}systems} where the base station is equipped with one-bit analog-to-digital/digital-to-analog converters (ADC/DACs). Considering training-based transmission, we assume the base station (BS) employs the  {\color{black}linear minimum mean-squared-error (LMMSE)} channel estimator and treats the channel estimate as the true channel to precode the data symbols. We derive {\color{black}an} expression {\color{black}for} the downlink achievable rate for matched-filter (MF) precoding. {\color{black}A detailed analysis of the resulting } power efficiency is pursued using our expression of the achievable rate. Numerical results are presented to verify our analysis. In particular it is shown that, compared with conventional massive MIMO systems, the performance loss in one-bit massive MIMO {\color{black}systems} can be compensated {\color{black}for by} deploying {\color{black}approximately} 2.5 times more antennas at the BS.
\end{abstract}

\begin{IEEEkeywords}
Massive MIMO, one-bit DACs, downlink rate, MF precoding.
\end{IEEEkeywords}

%
\IEEEpeerreviewmaketitle

\section{Introduction}
Massive MIMO is an emerging technology capable of scaling up {\color{black}the performance of} conventional MIMO by orders of magnitude. It {\color{black}has been shown} that, with {\color{black}a base station (BS) equipped with a very large number of antennas}, not only can the spectral efficiency and energy efficiency be significantly improved by employing simple linear signal processing techniques, but also the {\color{black}impact of imperfections} in the hardware implementation can be mitigated \cite{lu2014overview,emil2015massive}.

{\color{black}Most prior work has assumed that} each antenna element in the massive MIMO system is equipped with {\color{black}a costly high-resolution digital-to-analog converter (DAC), and hence has neglected} the nonlinear effect of the quantization. {\color{black}The cost of using high-resolution DACs} is manageable in conventional MIMO systems since the number of antennas is relatively small. However, for massive MIMO configurations employing large antenna arrays and many ADCs/DACs, the cost and power consumption will be prohibitive.
%

{\color{black}The use of one-bit quantizers has been proposed as a potential solution to this problem for some time \cite{nossek2006capacity,ivrlac2006}. However, there has been limited prior work evaluating the downlink performance of communication systems with one-bit DACs. 
Previous work has considered standard linear precoder designs and their performance in the context of low resolution DACs \cite{mezghani2009transmit} and in the context of massive MIMO with one-bit DACs \cite{ovais2016mmse,Saxena_2016_2}, showing satisfactory performance for small loading factors and well conditioned channels (e.g., i.i.d. Rayleigh). Non-linear \emph{Tomlinson-Harashima Precoding} has been considered in \cite{Mezghani_2008_G} for low resolution DACs showing still better performance than purely linear methods. In \cite{hela2016minimum} a nonlinear symbol-by-symbol vector optimization for one-bit DAC systems is proposed based on a $\ell_\infty$-norm relaxation of the discrete DAC output set and a minimum-distance criterion and shows that such precoding schemes significantly outperform linear precoders at the cost of an increased computational complexity. The authors in \cite{jacobsson2016quantized,Jacobsson_2016_2} successfully applied this approach to DACs with arbitrary resolution and higher order modulation using several different algorithms and compared the results to quantized linear methods, again observing similar performance gains. Recently, another nonlinear method  based on perturbation techniques has been proposed in \cite{Swindlehurst_2017}.  The derivation of achievable rates for multi-user systems with low resolution/one-bit DACs has also been considered in \cite{Kakkavas_2016} for standard MIMO and in \cite{jacobsson2016quantized} for massive MIMO implementations.}


In this paper, motivated by our recent work in \cite{yongzhi2016channel}, we consider a downlink massive MIMO system with one-bit DACs on each transmit antenna and derive a lower bound on the downlink achievable rate for matched-filter (MF) precoding. {\color{black} The key difference between our work and that cited above is that our derivation includes the effects of channel estimation error.} Based on the Bussgang decomposition, we first derive a closed-form expression for the downlink achievable rate for MF precoding, and then {\color{black}based on the obtained expression, we perform a detailed analysis of the system performance.} It is shown that, {\color{black}compared} with conventional massive MIMO, the performance loss due to the use of one-bit DACs {\color{black}systems} can be compensated {\color{black} for by} deploying approximately 2.5 times more antennas at the BS.

\section{System Model}
In this paper, we consider a downlink single-cell one-bit massive MIMO system with $K$ single-antenna terminals and an $M$-antenna BS. {\color{black}As depicted in Fig.~1, the BS is assumed to first apply an $M\times K$ linear precoder $\mathbf{W}$ to the vector $\mathbf{s}$ whose elements represent the symbols for each of the $K$ users. Then the DACs separately quantize the real and imaginary parts of the precoded signal using a single bit; i.e., only the sign of the real and imaginary part of the signal is retained.} Thus, the quantized {\color{black}transmit} signal can be expressed as
\begin{equation}\label{quantized_signal_orign}
\mathbf{y} = \mathcal{Q}\left(\mathbf{x}\right) = \mathcal{Q}\left(\mathbf{Ws}\right),
\end{equation}
where $\mathcal{Q}(.)$ is the one-bit quantization function, $\mathbf{x}\in\mathbb{C}^{M\times1}$ represents the precoded signal, and the data symbols $\mathbf{s}$ are assumed to satisfy $\E\{\mathbf{s}\mathbf{s}^H\} = \mathbf{I}$.
In this paper, in order to normalize the power of the output, we assume the {\color{black}quantized output falls in the set} $\mathcal{Y} = \frac{1}{\sqrt{2}}\{1+1j, 1-1j, -1+1j, -1-1j\}$. Then the received signal at the $K$ users is
\begin{align}\label{received_signal_origin}
\mathbf{r}_d &= \gamma\mathbf{H}^T\mathbf{y} + \mathbf{n}_d = \gamma\mathbf{H}^T\mathcal{Q}\left(\mathbf{Ws}\right) + \mathbf{n}_d,
\end{align}
where $\mathbf{H}$ is the $M\times K$ channel matrix between the $K$ users and the BS, $\mathbf{n}_d\sim\mathcal{CN}(\mathbf{0},\mathbf{I})$ is additive white Gaussian noise, and  $\gamma$ is a normalization parameter chosen to satisfy a long term total transmit power constraint $P_t$ at the BS, i.e., $\E\{\|\gamma\mathbf{y}\|_2^2\}=P_t$. Note that, owing to the one-bit DACs, the elements of the quantized analog signal $\mathbf{y}$ only have four {\color{black}states} in $\mathcal{Y}$, which implies $\E\{\|\mathbf{y}\|^2\}=M$. Therefore, we can obtain $\gamma = \sqrt{P_t/M}$.

\section{Uplink Channel Estimation and MF Precoding}
\subsection{Uplink Training}
\begin{figure}
  \centering
  \includegraphics[width=6cm]{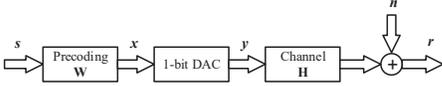}\\
  \vspace{-0.3cm}
  \caption{System architecture of the downlink one-bit massive MIMO system.}\label{Downlink_System_Archietecute}
  \vspace{-0.5cm}
\end{figure}
{\color{black}Assuming} training-based transmission, the channel matrix $\mathbf{H}$ is estimated at the BS in the uplink. We assume the $K$ users simultaneously transmit orthogonal pilot sequences to the BS, {\color{black}which we represent as} $\bm{\Phi}\in\mathbb{C}^{\tau\times K}$, and which thus satisfy $\bm{\Phi}^H\bm{\Phi} =\tau\mathbf{I}$. Therefore, the received training signal {\color{black}prior to quantization} at the BS is \cite{yongzhi2016channel}
\begin{align}\label{vec_received_signal_training}
\vect(\mathbf{Y}_p) &= \mathbf{y}_p = \vect\left(\sqrt{\rho_p}\mathbf{H}\bm{\Phi}^T + \mathbf{N}_p\right)\nonumber \\
&=\left(\bm{\Phi} \otimes \sqrt{\rho_p}\mathbf{I}_{M}\right)\underline{\mathbf{h}} + \underline{\mathbf{n}}_p,
\end{align}
where $\rho_p$ is the transmitted training power of each user, $\underline{\mathbf{h}} = {\rm vec}(\mathbf{H})$ and $\underline{\mathbf{n}}_p=\vect(\mathbf{N}_p)$.

Although we note that, unlike conventional MIMO systems, the assumption of $\tau=K$ is not in general optimal for one-bit MIMO \cite{yongzhi2016channel,yongzhi2016how}, in the sequel we will assume $\tau = K$ to simplify the analysis. We also note that although the one-bit quantization is a nonlinear operation, we can reformulate it as {\color{black}a statistically equivalent} linear operation using the Bussgang decomposition \cite{bussgang1952yq}. In particular, after the one-bit ADCs, the quantized uplink training signal can be reformulated as \cite{yongzhi2016channel}
\begin{align}\label{Training_Quantized_Signal}
  \mathbf{r}_p &= \mathcal{Q}(\mathbf{y}_p) = \mathcal{Q}\left(\left(\bm{\Phi} \otimes \sqrt{\rho_p}\mathbf{I}_{M}\right)\underline{\mathbf{h}} + \underline{\mathbf{n}}_p\right) \nonumber \\
  &= \tilde{\bm{\Phi}}{\underline{\mathbf{h}}} + \mathbf{A}_p{\underline{\mathbf{n}}}_p + \mathbf{q}_p,
\end{align}
where $\mathbf{r}_p\in\mathcal{Y}$ and $\tilde{\bm{\Phi}} = \mathbf{A}_p\left(\bm{\Phi} \otimes \sqrt{\rho_p}\mathbf{I}_{M}\right)$, $\mathbf{A}_p$ is the resulting Bussgang linear operator and $\mathbf{q}_p$ the statistically
equivalent quantization noise. {\color{black}Using the linear minimum mean-squared error (LMMSE) approach, the channel estimate is given by} \cite[Eq. (23)]{yongzhi2016channel}
\begin{equation}\label{channel_estimate}
{\underline{\hat{\mathbf{h}}}} =\tilde{\bm{\Phi}}^H\mathbf{r}_p
\end{equation}
with $\mathbf{A}_p=\alpha_p \mathbf{I}$ and $\alpha_p = \sqrt{2/(\pi(K\rho_p+1))}$.

Note that each element of $\hat{\mathbf{h}}$ can be expressed as a summation of random variables, i.e., $[\hat{\mathbf{h}}]_{n} = \sum_{i=1}^{MK}[\tilde{\bm{\Phi}}^H]_{n,i} r_{p,i}$. Although the channel estimate \eqref{channel_estimate} {\color{black}is in general} not Gaussian distributed due to the quantizer noise, we can approximate it as Gaussian according to Cram{\' e}r's central limit theorem \cite{Cramer2004random} assuming $K$ is sufficiently large. Therefore, in what follows we model each element of the channel estimate $\hat{\mathbf{h}}$ as Gaussian with zero mean and variance $\eta^2 = 2K\rho_p/\pi(1+K\rho_p)$.

\subsection{MF Precoding}
For the downlink transmission, we assume the BS considers the channel estimate as the true channel and employs matched-filter (MF) precoding to process the data symbols before broadcasting to the $K$ users. The MF precoding matrix is given by $\mathbf{W} = {\hat{\mathbf{H}}}^*$,
where we define inverse vectorization operator $\hat{\mathbf{H}} = {\rm unvec}({\hat{\underline{\mathbf{h}}}})$. Then according to the Bussgang decomposition, we reformulate the quantized signal $\mathbf{y}$ in~\eqref{quantized_signal_orign}~as
\begin{equation}\label{quantized_signal_MF}
\mathbf{y}_d =\mathcal{Q}({\hat{\mathbf{H}}}^*\mathbf{s})= \mathbf{A}_d{\hat{\mathbf{H}}}^*\mathbf{s} + \mathbf{q}_d,
\end{equation}
where the same definitions as in the previous sections apply, but replacing the subscript $p$ with $d$. The matrix $\mathbf{A}_d$ is
\begin{align}
\footnotesize
\mathbf{A}_d &= \sqrt{\frac{2}{\pi}}\diag(\mathbf{C}_{\mathbf{xx}})^{-\frac{1}{2}} = \sqrt{\frac{2}{\pi}}\diag\left({\hat{\mathbf{H}}}^*{\hat{\mathbf{H}}}^T\right)^{-\frac{1}{2}} \nonumber \\
& =\sqrt{\frac{2}{\pi K \eta^2}}\mathbf{I} \triangleq\alpha_d \mathbf{I},
\end{align}
where $\mathbf{C}_{\mathbf{xx}}$ is the auto-correlation matrix of $\mathbf{x}$.

\section{Downlink Achievable Rate Analysis and Performance Evaluation}
\subsection{Downlink Achievable Rate}
In this section, we {\color{black}derive} a lower bound on the downlink achievable rate for MF {\color{black}precoding}. Combining \eqref{received_signal_origin} and \eqref{quantized_signal_MF}, the received signal vector at the $K$ users is given by
\begin{equation}\label{received_signal}
\mathbf{r}_d = \gamma\mathbf{H}^T\left(\mathbf{A}_d{\hat{\mathbf{H}}}^*\mathbf{s} + \mathbf{q}_d\right) + \mathbf{n}_d.
\end{equation}
Thus, the received signal at the $k$th user can be expressed as
\begin{equation}\label{r_k}
r_{d,k} = \gamma\mathbf{h}_k^T\mathbf{A}_d{\hat{\mathbf{h}}}_k^*s_k + \gamma\mathbf{h}_k^T\mathbf{A}_d\sum_{i\neq k}^K{\hat{\mathbf{h}}}_i^*s_i + \gamma\mathbf{h}_k^T\mathbf{q}_d + {n}_{d,k},
\end{equation}
where the last three terms in \eqref{r_k} respectively correspond to
inter-user interference, quantization noise and AWGN noise.

Note that, owing to the nonlinear quantization of the one-bit DACs, the quantizer noise $\mathbf{q}_d$ is not distributed as Gaussian. However, we can obtain a lower bound on the achievable rate by making the worst-case assumption \cite{diggavi2001the} that in fact it is Gaussian with the same covariance matrix:
\begin{equation}
\mathbf{C}_{\mathbf{q}_d\mathbf{q}_d} = \mathbf{C}_{\mathbf{y}_d\mathbf{y}_d} - \mathbf{A}_d\mathbf{C}_{\mathbf{x}\mathbf{x}}\mathbf{A}_d^H.
\end{equation}
Thus, the ergodic achievable rate can be lower bounded by
\begin{equation}\label{ergodic_rate}
\footnotesize
{R_k} = \E\left\{{\log _2}\left( {1 + \frac{\gamma^2 |\mathbf{h}_k^T\mathbf{A}_d{\hat{\mathbf{h}}}_k^*|^2}{\gamma^2\sum_{i\neq k}^K|\mathbf{h}_k^T\mathbf{A}_d{\hat{\mathbf{h}}}_i^*|^2 + \gamma^2\mathbf{h}_k^T \mathbf{C}_{\mathbf{q}_d\mathbf{q}_d} \mathbf{h}_k^* + 1}} \right)\right\} .
\end{equation}

{\color{black} In order to obtain a closed-form expression for the ergodic achievable rate, we use the same technique as in \cite{jose2011pilot}:
we first rewrite the received signal of the $k$th user \eqref{r_k} as a known mean gain times the desired symbol, which depends on the channel distribution instead of the instantaneous channel, plus an effective {\color{black}noise term}:
\begin{equation}\label{hat_s_k}
\hat{s}_k = \E\left\{\gamma\mathbf{h}_k^T\mathbf{A}_d{\hat{\mathbf{h}}}_k^*\right\}s_k + \tilde{n}_{d,k},
\end{equation}
where $\tilde{n}_{d,k} $ is the effective noise
\begin{align}
{{\tilde n}}_{d,k} =& \left( \gamma\mathbf{h}_k^T\mathbf{A}_d{\hat{\mathbf{h}}}_k^* - \E\left\{\gamma\mathbf{h}_k^T\mathbf{A}_d{\hat{\mathbf{h}}}_k^*\right\} \right){s_k} \nonumber\\
&+  \gamma\mathbf{h}_k^T\mathbf{A}_d\sum_{i\neq k}^K{\hat{\mathbf{h}}}_i^*s_i + \gamma\mathbf{h}_k^T\mathbf{q}_d + {n}_{d,k}.
\end{align}
Next we define the linear minimum mean square error (LMMSE) estimate $\tilde{s}_k$ of $s_k$ based on $\hat{s}_k$
\begin{equation}
  \tilde{s}_k= \frac{\gamma \E\{\mathbf{h}_k^T\mathbf{A}_d{\hat{\mathbf{h}}}_k^*\} }{\gamma^2 |\E\{\mathbf{h}_k^T\mathbf{A}_d{\hat{\mathbf{h}}}_k^*\}|^2+ \E\{| {\tilde n}_{d,k}|^2\}} \hat{s}_k ,
\end{equation}
resulting in the following MSE:
\begin{equation}
\footnotesize
\E\{| s_k-\tilde{s}_k|^2\}=\E\{| \varepsilon_k |^2\}= \frac{\E\{| {\tilde n}_{d,k}|^2\}}{\gamma^2 |\E\{\mathbf{h}_k^T\mathbf{A}_d{\hat{\mathbf{h}}}_k^*\}|^2+ \E\{| {\tilde n}_{d,k}|^2\}} .
\end{equation}

Then, we can obtain a lower bound for the mutual information $I(s_k,\tilde{s}_k)$ with Gaussian input $s_k$ as
\begin{align}
I(s_k,\tilde{s}_k)&=  h(s_k)-h(s_k|\tilde{s}_k)
=  h(s_k)-h(s_k-\tilde{s}_k |\tilde{s}_k) \nonumber \\
&\geq  h(s_k)-h(\underbrace{s_k-\tilde{s}_k }_{\varepsilon_k})\geq \log_2\frac{1}{ \E\{| \varepsilon_k |^2\} }.\label{conditioning}
\end{align}
We obtain the first inequality in~\eqref{conditioning} since conditioning reduces entropy. The second inequality is due to the fact that $h({\varepsilon_k})$ is upper bounded by the entropy of a Gaussian random variable whose covariance is equal to the error variance $\E\{| \varepsilon_k |^2\}$ of the linear MMSE estimate of $s_k$. }
Therefore using this approach
a closed-form expression for the achievable rate can be obtained. Furthermore, substituting $\mathbf{A}_d = \alpha_d \mathbf{I}$ yields
\begin{equation}\label{Achievable_Rate}
R_k = \log_2 \left( 1 + \frac{\alpha_d^2\gamma^2\left| {\E\left\{ {\hat{\mathbf{{h}}}_k^T  {\mathbf{h}_k^*}} \right\}} \right|^2}{\alpha_d^2\gamma^2\Var\left({\hat{\mathbf{{h}}}_k^T{{\bf{h}}_k^*}}\right) + \textrm{UI}_k + \textrm{QN}_k +1}\right),
\end{equation}
where
\begin{equation}
\textrm{UI}_k = \alpha_d^2\gamma^2\sum_{i\neq k }^K \E\left\{\left|\hat{\mathbf{{h}}}_k^T \mathbf{h}_i^*\right|^2\right\}
\end{equation}
\begin{equation}
\textrm{QN}_k = \gamma^2(1-2/\pi)\E\left\{\left\|{\mathbf{{h}}}_k^T \right\|^2\right\}.
\end{equation}
Next we provide a closed-form expression for the achievable rate with MF precoding.

{\it Theorem 1}: For MF precoding, with imperfect CSI estimated by the LMMSE channel estimator, the downlink achievable rate of the $k$th user in a one-bit massive MIMO system {\color{black}is lower bounded by}
\begin{equation}\label{closed_rate}
R_k = \log_2\left(1+\frac{4M\rho_p P_t}{\pi^2(1+K\rho_p)(1+P_t)}\right).
\end{equation}

\begin{IEEEproof}
See Appendix A.
\end{IEEEproof}

\subsection{Performance Evaluation}

1) {\it Power Efficiency}: We first study power efficiency for the one-bit massive MIMO downlink.

{\it Case I}: If $\rho_p$ is fixed and $P_t = E_t/M$, where $E_t$ is fixed regardless of $M$, the downlink achievable rate converges to
\begin{equation}
R_k\rightarrow \log_2\left(1+\frac{4\rho_p E_t}{\pi^2(1+K\rho_p)}\right)
\end{equation}
as $M$ tends to infinity. We see that, although the BS {\color{black}is only equipped} with one-bit ADC/DACs, the total transmit power of the BS still can be reduced proportionally to $1/M$ while maintaining a given achievable rate when the channel estimation accuracy is fixed.

{\it Case II}: If $\rho_p = E_u/\sqrt{M}$ and $P_t = E_t/\sqrt{M}$, where $E_u$ and $E_t$ are fixed regardless of $M$, the downlink achievable rate converges to
\begin{equation}
R_k\rightarrow\log_2\left(1+\frac{4E_u E_t}{\pi^2}\right)
\end{equation}
when $M$ increases to infinity.
We see that the training power of the users and the total transmit power of the BS cannot be reduced as aggressively as in Case I where the {\color{black}accuracy of the} channel estimate is fixed. This is because when we {\color{black}reduce} the training power of the users, the channel {\color{black}estimation} accuracy will deteriorate. Therefore, we can only scale down $\rho_p$ and $P_t$ proportionally to $1/\sqrt{M}$ {\color{black}in order to maintain a given achievable rate}.

2) {\it {\color{black}Comparison} with conventional massive MIMO}: We next compare the downlink achievable rates between one-bit and conventional massive MIMO in terms of the number of antennas deployed at the BS. For the conventional massive MIMO system, we assume the BS employs perfect ADC/DACs with infinite resolution, which do not suffer from quantization loss. For this analysis, we denote the
number of antennas in the one-bit and conventional massive MIMO systems as $M_{\textrm{one}}$ and $M_{\textrm{conv}}$, respectively.
The downlink achievable rate in the conventional massive MIMO system is given by {\color{black}\cite{ngo2014multipair}}
\begin{equation}\label{typ_closed_rate}
R_{k,\textrm{conv}} = \log_2\left(1+\frac{M_{\textrm{conv}}\rho_p P_t}{(1+P_t)(1+K\rho_p)}\right).
\end{equation}
{\color{black} Comparing \eqref{closed_rate} with \eqref{typ_closed_rate}, we see that the terms inside the parentheses can be made equal by choosing $M=\pi^2 M_{\textrm{conv}} / 4 \approx 2.5 M_{\textrm{conv}}$. Thus, to achieve performance comparable to a conventional system, the one-bit system must deploy about 2.5 times more antennas.} {\color{black}We note that this ratio also holds for the zero-forcing precoder at low SNR as well.}

\section{Numerical Results}
For our simulations, we consider a single-cell one-bit massive MIMO downlink with $K = 10$ users.
\begin{figure}[!t]
  \centering
  \includegraphics[width=6cm]{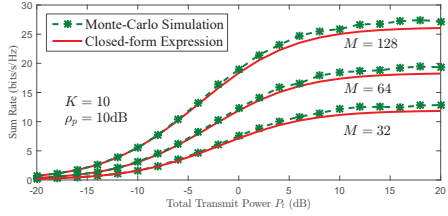}\\
  \vspace{-0.3cm}
  \caption{Sum Rate versus the total transmit power $P_t$ for different number of transmit antennas with $K = 10$ and $\rho_p = 10$dB.}\label{SumRate_MF}\vspace{-0.3cm}
\end{figure}
We first evaluate the validity of our closed-form expression for the achievable rate given in Theorem 1. Fig. \ref{SumRate_MF} shows the sum rate versus the total transmit power $P_t$ of the BS for different {\color{black}numbers} of transmit antennas $M=\{32,64,128\}$. The dashed lines represent the sum rate obtained numerically from \eqref{ergodic_rate}, and the solid lines are obtained by using the closed-form expression given in \eqref{closed_rate}. We see that the performance gaps between the Monte-Carlo results and the closed-form results are small. This indicates that our expression is a good predictor of the performance of the one-bit massive MIMO system.

\begin{figure}[!t]
  \centering
  \includegraphics[width=6cm]{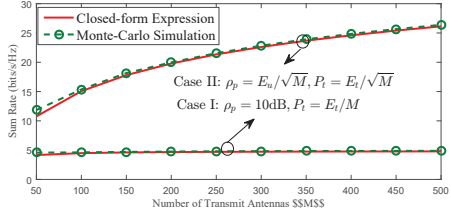}\\
  \vspace{-0.3cm}
  \caption{Sum rate versus the number of transmit antennas $M$ with $\rho_p=10$dB, $P_t = E_t/M$ in Case I, and $\rho_p = E_u/\sqrt{M}$, $P_t= E_t/\sqrt{M}$ in Case II.}\label{PowerEfficiency_MF}
  \vspace{-0.3cm}
\end{figure}
{\color{black}Next} we investigate the power efficiency {\color{black}of one-bit massive MIMO} for Case I and Case II. Fig. \ref{PowerEfficiency_MF} illustrates the sum rate versus the number of transmit antennas $M$ for MF precoding. In Case I, we assume $\rho_p = 10$dB is fixed and $P_t = E_t/M$, where $E_t = 10$dB. In Case II, we choose $\rho_p = E_u/\sqrt{M}$ and $P_t = E_t/\sqrt{M}$ where $E_u =E_t = 10$dB. As predicted in our analysis, the sum rates converge to a fixed constant in both cases.

\begin{figure}[!t]
  \centering
  \includegraphics[width=6cm]{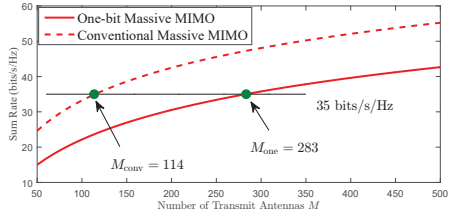}\\
  \vspace{-0.3cm}
  \caption{Comparison of the sum rate versus the number of transmit antennas between one-bit massive MIMO system and the conventional massive MIMO system with $\rho_p = P_t = 10$dB.}\label{Compare}
  \vspace{-0.5cm}
\end{figure}
Finally we compare the sum rates between the one-bit and conventional massive MIMO systems. Fig. \ref{Compare} shows the sum rate versus the number of transmit antennas with $\rho_p = P_t = 10$dB. {\color{black}The curves illustrate the fact that 2.5 more antennas are required by the one-bit system in order to achieve the same performance as the conventional system.} For example, in order to obtain the achievable rate of 35bits/s/Hz, $M_{\textrm{one}} = 283$ transmit antennas should be deployed in a one-bit massive MIMO system, {\color{black}compared with 114} for the conventional massive MIMO system. {\color{black}Thus we see how a large number of antennas can be used to compensate for loss of fidelity due to hardware imperfections}.

\section{Conclusions}
We considered a downlink massive MIMO system with one-bit DACs and derived a closed-form expression for the downlink achievable rate. Employing our obtained expression, we evaluated the power efficiency of such a system and showed that the total transmit power can be reduced by increasing the number of transmit antennas. Moreover, {\color{black}we} demonstrated that, with a matched-filter beamformer, the performance loss caused by the one-bit DACs can be compensated for by deploying {\color{black}approximately} 2.5 times more antennas at the BS, which confirms the benefit of the massive MIMO technique {\color{black}in overcoming hardware imperfections}.

\appendices
\section{}
According to Cram{\' e}r's central limit theorem, the elements of the channel estimate $\hat{\underline{\mathbf{h}}}$ can be approximated as Gaussian {\color{black}random variables} with zero mean and variance of $\eta^2$. Therefore,
\begin{equation}\label{E_h}
\E\{\hat{\mathbf{h}}_k^T\mathbf{h}_k^*\} = \E\{\|\hat{\mathbf{h}}_k^T\|_2^2\} = M\eta^2
\end{equation}
\begin{equation}
\Var\{\hat{\mathbf{h}}_k^T\mathbf{h}_k^*\} = \E\{|\hat{\mathbf{h}}_k^T\mathbf{h}_k^*|^2\} - \left(\E\{\hat{\mathbf{h}}_k^T\mathbf{h}_k^*\}\right)^2 = M\eta^2
\end{equation}
\begin{equation}\label{UI_k}
\textrm{UI}_k = \alpha_d^2\gamma M(K-1)\eta^2 \; .
\end{equation}
By substituting \eqref{E_h}-\eqref{UI_k} {\color{black}into \eqref{Achievable_Rate}}, Theorem 1 can be {\color{black}obtained}.

\bibliographystyle{IEEEtran}
\bibliography{reference}

\end{document}